\documentclass[twocolumn,amsmath,amssymb,11pt,superscriptaddress, nofootinbib]{revtex4-1}
\pdfoutput=1

\usepackage[english]{babel}
\usepackage{amssymb}
\usepackage{amsmath}
\usepackage{amsthm}
\usepackage[]{graphicx}
\usepackage{tensor}
\usepackage{color}
\usepackage{framed}
\usepackage{natbib}
\usepackage{dcolumn}
\usepackage{bm}
\usepackage{subcaption}

\usepackage{fancyhdr}
\usepackage[us,12hr]{datetime}

\graphicspath{
{./figures/}}

\usepackage{tikz}
\usetikzlibrary{decorations.pathmorphing, patterns,shapes}
\usetikzlibrary{calc}
\usetikzlibrary{intersections}
\usetikzlibrary{angles}
\usetikzlibrary{quotes}

\input epsf

\begin{document}

\allowdisplaybreaks
\begin{titlepage}

\title{Gravitational lensing of binary systems in wave optics}

\author{Job Feldbrugge}
\email{jfeldbrugge@perimeterinstitute.ca}
\affiliation{Perimeter Institute, 31 Caroline St N, Ontario, Canada}
\affiliation{Department of Physics, Carnegie Mellon University, 5000 Forbes Ave, Pittsburgh, PA 15217, USA}
\author{Neil Turok}
\email{nturok@perimeterinstitute.ca}
\affiliation{James Clerk Maxwell Building, University of Edinburgh, EH9 3FD, Edinburgh, Scotland}
\affiliation{Perimeter Institute, 31 Caroline St N, Ontario, Canada}

\begin{abstract}
We present the first detailed computations of wave optics effects in the gravitational lensing of binary systems. The field is conceptually rich, combining the caustic singularities produced in classical gravitational lensing with quantum (wave) interference effects. New techniques have enabled us to overcome previous barriers to computation. Recent developments in radio astronomy present observational opportunities which, while still futuristic, appear promising. 

\end{abstract}
\maketitle
\end{titlepage}


Einstein's theory of gravity underlies our understanding of the cosmos, and describes many remarkable phenomena including black holes, gravitational waves and gravitational lensing. Each is of great interest in its own right and has the potential to reveal new physics. In this Letter, we re-examine gravitational lensing in wave optics~\cite{Schneider, Nakamura:1999,Sugiyama:2019dgt}. Despite its ubiquity and universality, it has not yet been observed. Two recent developments have excited renewed attention~\cite{Katz:2019qug,Jow:2020rcy, Sammons:2020kyk,Wucknitz:2020spz}. Gravitational microlensing studies have revealed thousands of lenses including many exoplanets~\cite{Mao:2012za,Herrera_Mart_n_2020}. Fast Radio Bursts (FRBs) and pulsars are being detected in increasing numbers~\cite{Petroff:2019tty}. These coherent, pointlike sources could allow us to observe interference effects in gravitational lensing for the first time. 

The gravitational lensing of light from a distant source by a point mass $M$ is characterized by the Einstein angle, $\theta_E = \sqrt{2 R_S/d}$, with $R_S= 2 G M/c^2$ the Schwarzchild radius and $d$ the lens-observer distance~\cite{Einstein}. A ray optics description applies for wavelengths $\lambda$ much smaller than any relevant scale. Consider two initially parallel light rays. Their separation at the lens, $d\, \theta_E$, may be treated as that of a Young's double slit experiment. The fringe separation $\lambda/\theta_E$ is of order the classical deflection $d\,\theta_E$,  for $\lambda \sim R_S$. Equally, $\lambda$ is then of order the path difference $\sim R_S$. For lens masses from planets to stars, $R_S$ ranges from millimeters to kilometers so one might hope to observe wave optics effects with radio telescopes at these wavelengths. 

To observe the interference fringes, the observational aperture must be {\it smaller} than the fringe separation $\lambda/\theta_E$. This is  the condition that the two lensed images are {\it not} resolved. The situation is like Young's experiment, where one cannot both see the interference fringes and tell which slit the photon passed through. Remarkably, in the lensing of coherent point sources, like pulsars and FRBs, such effects operate over galactic, or even cosmological, scales. 

Interference patterns like those relevant here are surprisingly hard to compute. The required Fresnel-Kirchhoff  integrals are highly oscillatory and their convergence is delicate. This difficulty has thwarted detailed calculations in all but the simplest case, of lensing by a single, isolated point mass~\cite{Nakamura:1999}. Unfortunately, that case is too simple, lacking the intricate caustic structure of generic examples. Nor is it typical observationally, since most stars and planets occur in bound systems. Recently, we have overcome this barrier by developing  an efficient generalization of stationary phase methods, exploiting Picard-Lefschetz theory~\cite{Feldbrugge:2019fjs}. The idea is to flow the original, real integration contour onto a set of complex ``Lefschetz thimbles,'' each associated to a saddle point. Cauchy's theorem ensures the original integral is equivalent to a sum of absolutely convergent integrals. In Ref.~\cite{Feldbrugge:2019fjs}  we considered lensing phases meromorphic in the lens plane coordinates. Here, we generalize this to include logarithmic singularities. Whilst our main focus is on theoretical aspects, at the end we comment briefly on observational prospects. 

\begin{figure}
\centering
\resizebox {0.39\textwidth}{!}{
\begin{tikzpicture}[rotate=-90]
\node at (2.4,0.73)     [right] {$r_{lo}$};
\node at (2.22,2.75) [left]  {$r_{os}$};

\filldraw[black] (1.7,2.4) coordinate (xi)       circle (1pt) node[below left]  {}
                 (1.3,5) coordinate (eta)      circle (1pt) node[below left]  {} node[above] {source}
                 (3.5,3) coordinate (lens)     circle (1pt) node[above right] {lens}
                 (3.5,0) coordinate (observer) circle (1pt) node[below]       {observer};

\draw (observer) -- (xi) -- (eta);
\draw[dotted] (3.5, 3.) -- (observer)
              (observer) -- (eta);
\draw pic["$\bm{\theta}_{ls}$", draw=black, <->, angle eccentricity=1.2, angle radius=1.4cm] {angle=lens--observer--eta}
      pic["$\bm{\theta}$", draw=black, <->, angle eccentricity=1.1, angle radius=2.0cm] {angle=lens--observer--xi};

\draw[black] ([shift=(75.5:3cm)]3.5,0) arc (75.5:140:3cm);
\end{tikzpicture}
}
\caption{Geometry of interfering paths. }
\label{fig:LensSetup}
\end{figure}
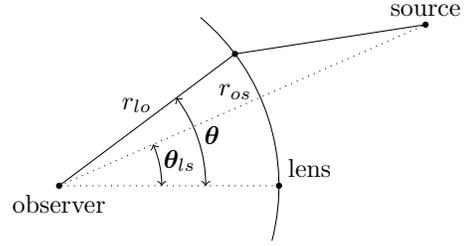

The gravitational lensing of coherent, monochromatic radiation with angular frequency $\omega$ may be treated quantum mechanically, as a path integral for a photon of fixed energy $E=\hbar \omega$. Consider a perturbed spacetime, with $\mathrm{d}s^2 = -(1+2U(\bm{r}))\mathrm{d}t^2 + (1-2U(\bm{r}))\mathrm{d}\bm{r}^2$, $U(\bm{r})$ being the gravitational potential (in units where $c=1$). The worldline action for a particle of mass $m$ is $-m \int ds$. For fixed initial energy $E$ we must add a boundary term $+E(t_f-t_i)$, with $t_i$ and $t_f$ the initial and final times~\cite{Feldbrugge:2019fjs}. The canonical momentum $\bm{p}$ obeys $\bm{p}^2/(1-2 U)+m^2 =E^2/(1+2 U)$. Taking $m\rightarrow 0$, to first order in $U$ the action becomes
\begin{align}
{\cal S}=\int \bm{p}\cdot \bm{dr}\approx E \int |\bm{dr}| \left(1-2 U(\bm{r})\right).
\label{e1}
\end{align}
 The quantum amplitude we seek is an integral over all paths from the source to the observer, weighted by $e^{ i{\cal S}/\hbar}$. For small angles, and where the particle's deflection occurs over a relatively small region, one can use the thin lens approximation. To first order, there are two contributions to (\ref{e1}). First, the Euclidean length of the lensed path (see Fig.~\ref{fig:LensSetup}) exceeds that of a straight line from source to observer by $\approx {1\over 2} |\bm{\theta}-\bm{\theta}_{ls}|^2r_{lo}r_{os}/(r_{os}-r_{lo})$. Second, the line integral of the potential $\int d r U(\bm{r})$, with $U=-G M_i/|\bm{r}-\bm{r}_i|$ for a mass $M_i$ at $\bm{r}_i$ is, for small $|\bm{\theta}- \bm{\theta}_i|$, approximated by $2G M_i \ln |\bm{\theta}- \bm{\theta}_i|$, up to a constant.  Expressing angles in terms of the Einstein angle, $\bm{\theta}\equiv \theta_E \bm{x}$ and $\bm{\theta}_{ls}\equiv \theta_E \bm{y}$, with $\theta_E\equiv \sqrt{4 G M (r_{os}-r_{lo})/(r_{lo} r_{os})}$ and $M$ the total mass, the path integral amplitude, normalized to unity in the absence of a lens, becomes
  \begin{align}
\Psi[\bm{y}] = \frac{\Omega}{2\pi i} \int_{\bm{x}}\,e^{i \Omega \left(
\frac{1}{2}|\bm{x}-\bm{y}|^2 +\varphi(\bm{x})\right)}.
\label{e2}
\end{align}
Here, $\Omega \equiv 4 G M \omega$, $\varphi(\bm{x})\equiv -\sum_i f_i \log |\bm{x}-\bm{x}_i|$, $f_i$ is the mass fraction and $\bm{x}_i$ the location of each lensing mass and  $\int_{\bm{x}} \equiv \int \mathrm{d}^2\bm{x}$ is the integral over the lens. Even a weak gravitational potential can have a large, nonperturbative effect, represented by nontrivial saddles in (\ref{e2}), since the lensing effect accumulates over large distances. Above, the angular positions of the lens masses were assumed fixed. If, instead, they drift together across the sky, $\bm{x}_i \rightarrow \bm{x}_i +\bm{\mu} t$, the amplitude is  translated:  $\Psi[\bm{y}] \rightarrow \Psi[\bm{y}-\bm{\mu} t]$.

Imagine restricting the integral over the lens to an area $A$, {\it i.e.}, masking the area outside $A$. The integrated intensity, $\int_{\bm{y}} |\Psi[\bm{y}]|^2=A$, independent of $\varphi(\bm{x})$. This is {\it unitarity}, which ensures that lensing preserves the total energy flux.  
  
At large $\Omega$, ray optics applies. Multiplying (\ref{e2}) by its complex conjugate, with $\bm{x}\rightarrow \bm{x}'$, and setting $\bm{ u}={1\over 2}(\bm{ x}'+ \bm{ x})$,  $\bm{\Delta}={1\over 2}(\bm{x}'- \bm{ x})$ gives
   \begin{align}
{\Omega^2\over 4 \pi^2} \int_{\bm{u},\bm{\Delta}} \,e^{-i \Omega \left(\varphi(\bm{u}+\bm{\Delta})-\varphi(\bm{u}-\bm{\Delta})+2 \bm{\Delta}\cdot(\bm{u-y})\right)}
\label{e3}
\end{align}
for the intensity $I$. Taylor expanding the exponent in $\bm{\Delta}$ and rescaling $\bm{\Delta}\rightarrow \bm{\Delta}/\Omega$ leaves the linear, $\Omega$-independent term dominant at large $\Omega$. Relabelling $\bm{u}$ as $\bm{x}$, we obtain
  \begin{align}
 I(\bm{y})\approx  \int_{\bm{x}}  \delta\left(\bm{\nabla}\varphi(\bm{x}) +\bm{x} -\bm{y}\right) =\sum_s {1\over |D(\bm{x}_s(\bm{y}))|}\,,
\label{e4}
\end{align}
where $D(\bm{x})=\det(\delta_{ij}+\partial_i\partial_j \varphi(\bm{x})),$ $i,j=1,2,$ and derivatives are taken wrt the indicated argument. The sum runs over solutions $\bm{x}_s(\bm{y})$ to
  \begin{align}
\bm{\nabla}\varphi(\bm{x}) +\bm{x} -\bm{y}=0,
\label{e5}
\end{align}
{\it i.e.}, the saddle point equation for the amplitude (\ref{e2}).
Although we derived (\ref{e4}) in an approximation, for large $\Omega$,  it respects unitarity as is seen by integrating the left hand side over all $\bm{y}$. 

To leading order in $\varphi$, (\ref{e5}) gives $\bm{x} \approx \bm{y}$. Then $D(\bm{y})\approx 1+\bm{\nabla}^2 \varphi(\bm{y})$ and Eq.~(\ref{e4}) yields a Poisson-like equation $-\bm{\nabla}^2 \varphi(\bm{y})\approx I(\bm{y})-1$. Hence, in the high frequency, weak lensing regime one can directly retrieve the lensing phase from the intensity. For gravitational lensing, there is a further simplification. Namely, if $I(\bm{y})-1$ is approximated by a sum of delta functions, chosen to match its multipole moments, the solution for $\varphi$, being a sum of logarithms, will closely approximate the exact, nonlinear solution. In the large $\Omega$ limit, equations (\ref{e4}) and (\ref{e5}) form the basis of an iterative retrieval algorithm for $\varphi$. Later, we shall discuss how  $\Omega$ may also be retrieved. 

\begin{figure}
\centering
\begin{subfigure}[b]{0.24\textwidth}
\includegraphics[width=\textwidth]{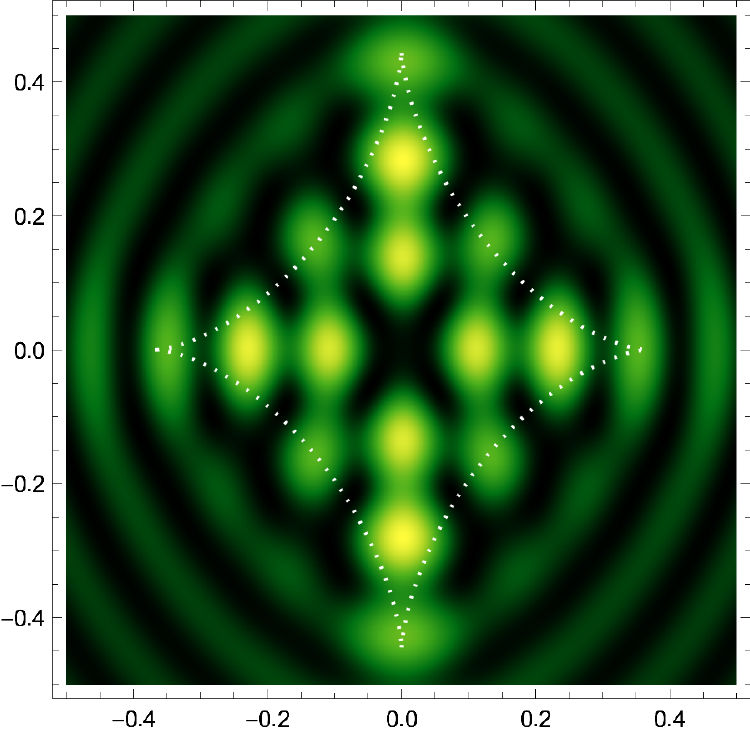}
\caption{$\Omega=25$}
\end{subfigure}~
\begin{subfigure}[b]{0.24\textwidth}
\includegraphics[width=\textwidth]{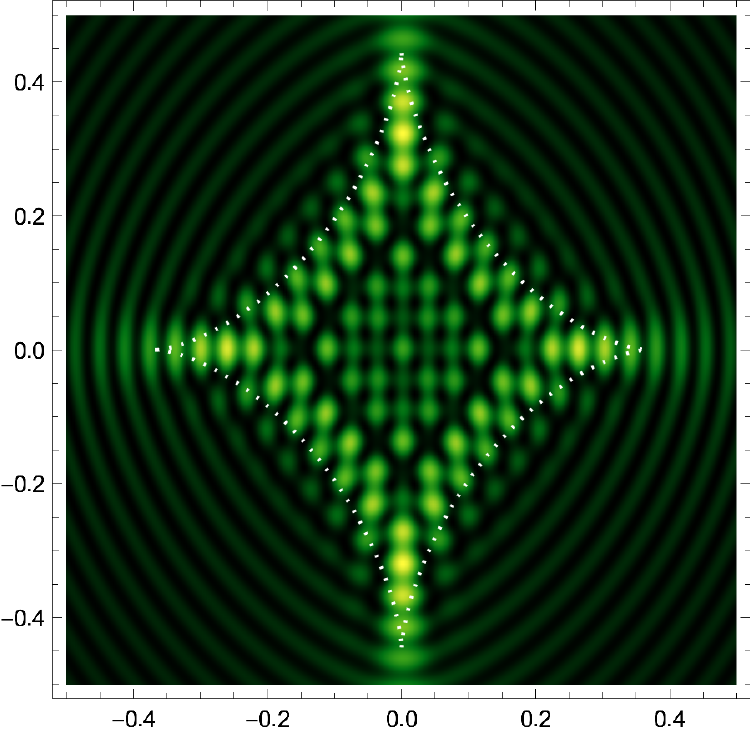}
\caption{$\Omega=75$}
\end{subfigure}
\caption{Single lens with shear $\gamma=0.2$. Caustic curve shown dotted in white.}
\label{fig:Shear}
\end{figure}

 
For a single gravitational lens, $\varphi(\bm{x})=-\ln x$ and $D(\bm{x})=1-1/x^4$.   Eq.~(\ref{e5}) has two solutions, $\bm{x}_{\pm}={\bm{y}\over  2 y} (y\pm \sqrt{4+y^2})$, corresponding to two classical rays. One passes through $\bm{x}_+$, on the same side as the source and outside the Einstein ring $x=1$. This ray dominates the intensity in the weak lensing regime. The other ray passes through $\bm{x}_-=-\bm{x}_+/x_+^2$,  on the opposite side of the lens and inside the Einstein ring.  Inserting these into (\ref{e4}), we obtain $I=(2+y^2)/(y\sqrt{4+y^2})$. In wave optics, (\ref{e2}) yields $I={\pi \Omega\over 1-e^{-\pi \Omega}} \left| _1 F_1\left(i \Omega/2, 1; i \Omega \,y^2/2\right)\right|^2$, with $_1 F_1$ a confluent hypergeometric function, which approaches the ray optics result at large $\Omega$~\cite{Schneider}. 

As a step towards greater realism, we add a background, shear potential so $\varphi(\bm{x}) = -\ln|\bm{x}| -{1\over 2} \gamma (x_1^2-x_2^2)$~\cite{Nakamura:1999}. The additional term could, for example, represent the tidal force from a second mass. There are four saddle point solutions to (\ref{e5}), with two or four being real depending on whether $\bm{y}$ is outside or inside the caustic curve (the locus in $\bm{y}$ where $D(\bm{x})$ vanishes). We perform the lensing integral (\ref{e2}) by passing to polar coordinates $\bm{x}=(x_1,x_2)=r(\cos \theta,\sin \theta)$. Only the radial integral is subtle because it involves an infinite number of oscillations. We flow the $r$ contour onto the relevant Lefschetz thimbles, and then integrate over $r$ and $\theta$ numerically. Typical intensity maps are shown in Fig.~\ref{fig:Shear}.

\begin{figure}
\centering
\begin{subfigure}[b]{0.23\textwidth}
\includegraphics[width=\textwidth] {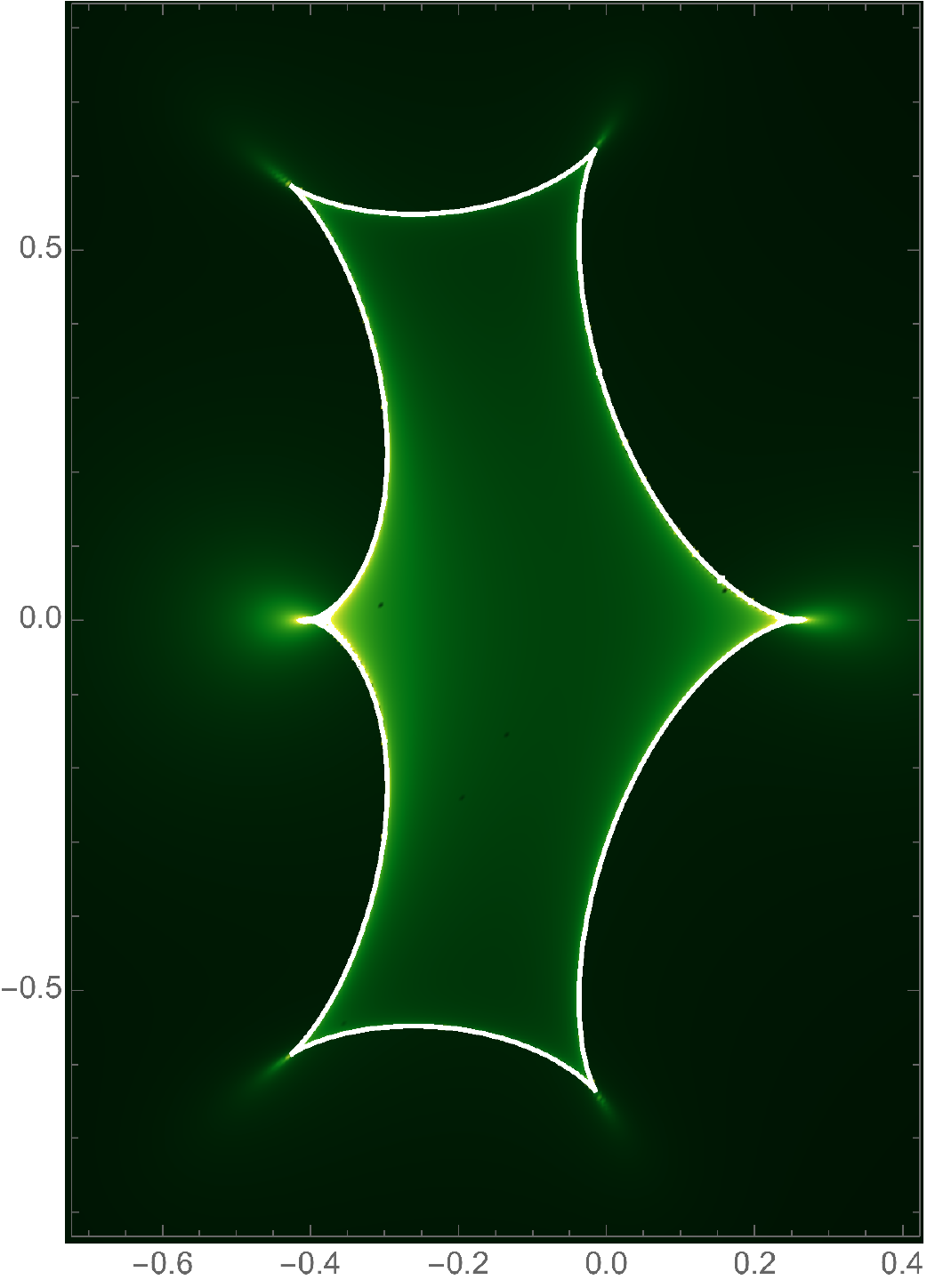}
\caption{ray optics}
\end{subfigure}~
\begin{subfigure}[b]{0.23\textwidth}
\includegraphics[width=\textwidth]{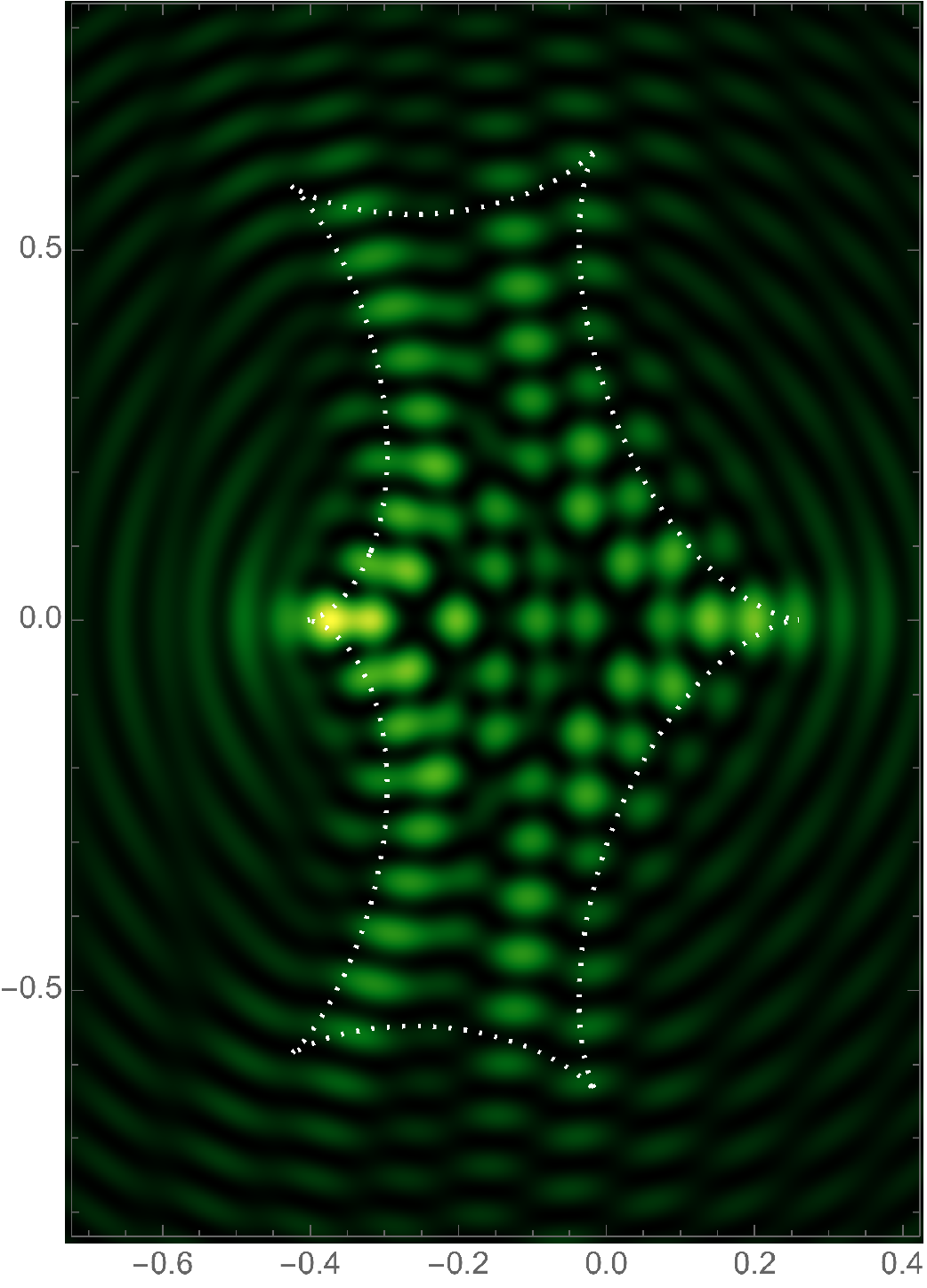}
\caption{wave optics ($\Omega=50$)}
\end{subfigure}
\caption{Ray and wave optics intensities for binary lens with $f_1=2 f_2$ and $a=1/2$}\label{fig:DoubleRayWave}
\end{figure}

For a binary lens,  we take $\varphi(\bm{x}) = -f_1 \log|\bm{x}+\bm{r}| -f_2 \log| \bm{x} - \bm{r}|$, with $f_1+f_2=1$, $f_1>f_2$ and $\bm{r}=(a,0)$. As an example, we take $f_1=2 f_2$, $a={1\over 2}$. The caustic curve ${\cal C}$ is shown in white in Fig.~\ref{fig:DoubleRayWave}(a). For each $\bm{y}$, Eq.~(\ref{e5}) has five roots for $\bm{x}$. All are real for $\bm{y}$ inside ${\cal C}$. As $\bm{y}$ approaches ${\cal C}$ from within, we encounter a {\it fold} singularity, where two saddles merge, or a {\it cusp} singularity, where three saddles merge. At these singularities, the ray optics intensity diverges. As $\bm{y}$ passes outside ${\cal C}$, the intensity drops, very sharply at a fold and as a ``whisker" with $I\propto (\Delta y)^{-1}$ beyond a cusp. The leftmost and rightmost cusps are the brightest. 


For a wave optics treatment, we use elliptic polar coordinates with foci at $\pm \bm{r}$, $\bm{x}(\tau,\sigma)=  a (\cosh \tau \cos \sigma, \sinh \tau \sin \sigma)\,$, with $0<\tau<\infty$ and $0<\sigma\leq 2 \pi$. At each $\sigma$, we  flow the $\tau$ contour onto the relevant Picard-Lefschetz thimbles, before integrating over $\tau$ and $\sigma$ numerically. It is convenient to extend the $\tau$ integral to the real line before analytic continuation. It becomes
\begin{align}
\int_{-\infty}^{\infty}  \mathrm{d}\tau e^{i \Omega\left({1\over 2} (\bm{x}(\tau,\sigma)-\bm{y})^2+\varphi(\tau,\sigma)\right)+\ln J(\sigma,\tau)}\,,
\label{ej1}
\end{align}
with $\varphi(\tau\,,\sigma)=-f_1\log(a C_+) - f_2 \log(a C_-)$, $C_{\pm}=\cosh\tau \pm \cos \sigma$, $J(\tau,\sigma)= {a^2\over 2}(\cosh 2\tau -\cos 2\sigma)$. At real $\sigma$, the integrand is periodic in ${\rm Im}(\tau$) with period $2\pi$, with five complex saddle points in the strip $-\pi \leq \text{Im}[\tau] < \pi$. For each $\sigma$, there are branch points at $\tau=\pm i \sigma$, $\pm i (\sigma-\pi)$ and $\pm i(\sigma -2 \pi)$, where the exponent in (\ref{ej1}) diverges.  We take the branch cuts to run away from the origin. When the integrand is analytically continued, the integral has two wedges of convergence, which asymptote to $\tau = -\infty - i\frac{\pi}{4}$ and $\tau = +\infty + i\frac{\pi}{4}$. The relevant Picard-Lefschetz thimbles are found by flowing the $\tau$ contour. The final intensity is shown in Fig.~\ref{fig:DoubleRayWave}. Near the leftmost and rightmost cusps, the pattern resembles that for a point lens in a shear field.  For further details and a selection of intensity maps, see \cite{website}.


The lensing amplitude (\ref{e2}) has the following remarkable mathematical property. It may be viewed as a unitary transformation $(\Omega/ 2 \pi i) \int_{\bm{x}} e^{i {1\over 2} \Omega (\bm{y}-\bm{x})^2}$ of the phase $e^{i\Omega \varphi(\bm{x})}$. The inverse transformation is $i \int_{\bm{y}}e^{-i {1\over 2} \Omega (\bm{z}-\bm{y})^2}$. Applying the latter to $\Psi[\bm{y}]$ yields $e^{i\Omega \varphi(\bm{z})}$, with an exactly uniform intensity for {\it any} $\varphi(\bm{z})$. Hence $\Omega$ may be directly recovered from the lensing amplitude. Consider a single point lens first. Perform the inverse transformation at $\Omega'=\Omega+\Delta$, with $\Delta>0$. One obtains the single lens pattern scaled by $\sqrt{\Delta/\Omega}$. As $\Delta\downarrow 0$,  the intensity $I$ approaches unity everywhere except for a spike at the origin. For $\Delta<0$, one conversely obtains a {\it hole} in the intensity, effectively due to a negative mass lens, whose size shrinks to zero as $\Delta\uparrow 0$. Applied to a multiple point lens, one obtains the corresponding spikes and holes, which disappear at $\Delta=0$. In principle, lens masses and positions can be recovered with extreme accuracy~\cite{FTprep}.

Finally, let us turn to observational prospects (see also \cite{Katz:2019qug,Jow:2020rcy,Sammons:2020kyk,Wucknitz:2020spz}). Optical telescopes currently detect around 2000 gravitational microlensing events annually, due to stars in our galaxy. The optical depth (fraction of the sky lensed) by these stars is around $10^{-6}$. With large numbers of radio sources - potentially, $10^4$ or more FRB's per day and up to $10^5$ pulsars, gravitational lensing events may be seen. In handy units $\theta_E\approx 1 \,{\rm mas} (M/M_\odot)^{1\over 2} (8 \,{\rm kpc}/d)^{1\over 2} $ and $\Omega\approx 1.2\times 10^{5} (M/M_\odot) (\nu/{\rm GHz})$, from which the fringe separation angle  $\theta_F = \pi \theta_E/\Omega \approx 0.03 \, \mu{\rm as} (M_\odot/M)^{1\over 2} (8 \,{\rm kpc}/d)^{1\over 2}  ({\rm GHz} /\nu)$. For typical lensing stars in the galactic bulge,  $M\sim 0.3 M_\odot$ and $d\sim 8$ kpc. Typical angular speeds are $\mu\sim 5\, {\rm mas/yr} $ so microlensing events have timescales of months~\cite{Mao:2012za}, with exoplanets seen as ``bumps" on the light curves of their stars~\cite{Mao:2012za,Herrera_Mart_n_2020}. Diffraction fringes would cross a radio telescope in a time $\theta_F/\mu \sim$ minutes to days.  FRB's have so far only been detected at low frequencies, where telescopes have wider fields of view. However, they have fairly flat spectra up to 8 GHz~\cite{Gajjar:2018bth}. If they can be localized using outrigger telescopes at GHz frequencies, a set of higher frequency telescopes pointed at them simultaneously could together observe fringes spaced by $\approx 300 ({\rm 100 GHz} /\nu)$ km on the Earth's surface. The $\nu$-dependence would be a key signature of wave optical gravitational lensing.  

\noindent {\it Acknowledgements:}
We thank D. Jow and U-L. Pen for interesting us in this problem.
Research at Perimeter Institute is supported in part by the Government of Canada through the Department of Innovation, Science and Economic Development Canada and by the Province of Ontario through the Ministry of Colleges and Universities.
\bibliographystyle{utphys}
\bibliography{library}

\end{document}